\documentclass[twocolumn,prl,showpacs,preprintnumbers,amsmath,amssymb]{revtex4}
\usepackage{graphicx}
\usepackage{dcolumn}
\usepackage{bm}

\font\scripti=cmmi7
\font\scriptscripti=cmmi5
\def\sib#1{\setbox0 = \hbox{\scripti #1}
  \kern-.02em\copy0\kern-\wd0
  \kern.04em\box0} 
\def\ssib#1{\setbox0 = \hbox{\scriptscripti #1}
  \kern-.02em\copy0\kern-\wd0
  \kern.04em\box0} 
\font\tenib=cmmib10 
\skewchar\tenib='177 \skewchar\tenib='177 \skewchar\tenib='177
\def\pbold#1{\setbox0 = \hbox{$ #1 $}
  \kern-.022em\copy0\kern-\wd0
  \kern.011em\copy0\kern-\wd0
  \kern.011em\copy0\kern-\wd0
  \kern.011em\copy0\kern-\wd0
  \kern.011em\box0} 

\usepackage{graphicx}
\usepackage{dcolumn}
\usepackage{bm}
\usepackage{color}

\def\up{\uparrow}
\def\dwn{\downarrow}

\def\lesssim{\ \raise.3ex\hbox{$<$}\kern-0.8em\lower.7ex\hbox{$\sim$}\ }
\def\gesim{\ \raise.3ex\hbox{$>$}\kern-0.8em\lower.7ex\hbox{$\sim$}\ }

\begin{document}
\title{Non-Hermitian Ferromagnetism in an Ultracold Fermi Gas}
\author{Hiroyuki Tajima}
\author{Kei Iida}
\affiliation{Department of Mathematics and Physics, Kochi University, Kochi 780-8520, Japan}
\date{\today}
\begin{abstract}
We develop a non-Hermitian effective theory for a repulsively interacting Fermi 
gas in the excited branch.  The on-shell $T$-matrix is employed as a complex-valued 
interaction term, which describes a repulsive interaction between atoms in the 
excited branch and a two-body inelastic decay to the attractive branch.
To see the feature of this model, we have addressed, in the weak coupling regime,
the excitation properties of a repulsive Fermi polaron as well as the time-dependent 
number density. 
The analytic expressions obtained for these quantities qualitatively show a good 
agreement with recent experiments.
By calculating the dynamical transverse spin susceptibility
in the random phase approximation, we show that a ferromagnetic system 
with nonzero polarization undergoes a dynamical instability 
and tends towards a heterogeneous phase. 
\end{abstract}
\maketitle
{\it Introduction}---
An ultracold atomic gas has attracted much attention in modern physics
because various quantum many-body phenomena manifest themselves
due to controllability of such physical parameters as 
scattering length and density~\cite{Bloch,Giorgini}. For example,
the realization of crossover from the Bardeen-Cooper-Schrieffer (BCS) 
Fermi superfluid to the molecular Bose-Einstein condensation (BEC)
or vice versa~\cite{Regal,Zwierlein} by tuning the attractive interaction 
via a Feshbach resonance~\cite{Chin} has opened a new frontier for 
studies of strongly correlated quantum systems~\cite{Zwerger,Strinati,Ohashi}. 
Nowadays, the ultracold atomic gas acts as a quantum simulator 
in various research fields ranging from condensed matter to nuclear 
physics~\cite{Strinati,Ohashi,Gezerlis,vanWyk,Horikoshi1,Horikoshi2}.
\par
In cold atomic physics, what kind of many-body state of repulsively 
interacting Fermi gases occurs is a long-standing problem~\cite{Massignan}.
A ferromagnetic phase transition is expected to occur
due to the intrinsically short-range repulsion, given the analogy
with the so-called Stoner model for many-electron systems~\cite{Stoner}.
While a repulsive two-spin-component Fermi gas with a positive scattering 
length $a$ between different spin components of the same mass $m$ has been 
realized by using a Feshbach resonance~\cite{Jo}, however, it suffers 
a two-body inelastic decay to the molecular ground state~\cite{Pekker,Sanner}.
In fact, the positive scattering length in this system 
is accompanied by the two-body bound state whose binding energy 
is given by $1/ma^2$.  Although the repulsive Fermi gas can thus be 
realized only as a metastable state denoted by the excited or 
repulsive branch, the ferromagnetic phase is known
to occur in a magnetic domain-wall configuration under 
such a non-equilibrium condition~\cite{Valtolina}. 
Furthermore, a quasiparticle excitation in the spin-polarized 
limit, that is, a repulsive Fermi polaron, has intensively been studied in 
recent experiments~\cite{Kohstall,Koschorreck,Scazza}.
Most recently, the out-of-equilibrium dynamics of repulsive 
Fermi gases has been experimentally explored, which shows
that the metastable ferromagnetic state does not simply decay into 
the thermal equilibrium state in the attractive branch~\cite{Amico}, 
but a heterogeneous phase can appear by involving 
micro-scale phase separation between the states in the repulsive and 
attractive branches~\cite{Scazza:2020}. 
\par
To understand the ferromagnetic state in this system, various 
theoretical efforts beyond the mean-field approximation
such as quantum Monte-Carlo simulation~\cite{Pilati}, 
perturbation theories~\cite{Duine,He} and non-perturbative approaches~\cite{He3} have been made.
Effects of nonzero effective range~\cite{He2,Massignan2} and 
trap potentials~\cite{Sogo,Sandri} have also been investigated. 
Several dynamical properties such as spin drag relaxation~\cite{Duine2}, 
spin-dipole modes~\cite{Recati}, 
dynamical spin response~\cite{Sodemann,Sandri,Mistakidis}, and 
dynamical instability of spin spiral~\cite{Conduit} have been pointed out.
That being said, a theory for {\it non-equilibrium} ferromagnetism 
has not been well established.
In such a case, the Hamiltonian involves a non-Hermitian term 
associated with the two-body inelastic decay.  Recently,
non-equilibrium effects have theoretically been discussed in a dissipative Fermi-Hubbard model~\cite{Nakagawa,Cui2}.
For the case of an attractive interaction, a non-Hermitian BCS superfluid state with the two-body inelastic decay has 
been studied~\cite{Yamamoto,Iskin}.
Also, the non-equilibrium 
BCS-BEC crossover in a driven-dissipative Fermi gas is 
under investigation~\cite{Kawamura}.  
We remark that 
in nuclear physics, empirical optical potential models, which
involve non-Hermitian potential terms, are often used for 
simplified description of direct reactions~\cite{Feshbach,Varner,Hodgson,Brandan,Muga}.
\par
In this work, we develop a non-Hermitian effective theory 
for repulsively interacting two-spin-component Fermi 
gases in the excited branch.  We model the non-Hermitian 
interaction for the inelastic two-body decay
from the on-shell two-body $T$-matrix 
at finite momenta.  To see how to connect between 
the complex-valued interaction and the existing
experimental data, we derive excitation properties 
of a repulsive Fermi polaron in the polarized limit
and also time-dependent number density 
in the unpolarized case.  We show that the dynamical 
transverse spin susceptibility calculated within 
the random phase approximation (RPA) exhibits a dynamical 
instability in the long wave-length limit due to the 
non-Hermitian interaction in a metastable ferromagnetic state. 
\par
{\it Non-Hermitian effective Hamiltonian}---
For a two-spin-component Fermi gas with 
short-range interactions, the Pauli principle allows us to 
write the effective Hamiltonian as
\begin{eqnarray}
H_{\rm eff}&=&\sum_{\bm{p},\sigma}\xi_{\bm{p},\sigma}c_{\bm{p},\sigma}^\dag c_{\bm{p},\sigma}
+\sum_{\bm{k},\bm{k}',\bm{P}}U_{\rm R}(\bm{k},\bm{k}',\bm{P})\cr
&&\times c_{\bm{P}+\bm{k}'/2,\up}^\dag c_{\bm{P}-\bm{k}'/2,\dwn}^\dag
c_{\bm{P}-\bm{k}/2,\dwn}c_{\bm{P}+\bm{k}/2,\up},
\label{Heff}
\end{eqnarray}
where $\xi_{\bm{p},\sigma}=\varepsilon_{\bm{p}}-\mu_{\sigma}\equiv p^2/2m-\mu_{\sigma}$ 
is the kinetic energy of a Fermi atom with momentum $\bm{p}$, 
pseudospin $\sigma=\up,\dwn$, and mass $m$ 
measured from the chemical potential $\mu_{\sigma}$, and
$c_{\bm{p},\sigma}$ is the annihilation operator. 
$U_{\rm R}(\bm{k},\bm{k}',\bm{P})$ is the $\up$--$\dwn$
repulsive interaction that will be specified below as 
function of the relative momenta $\bm{k}$ and $\bm{k}'$
of incoming and outgoing two particles,  as well as
the center of mass momentum $\bm{P}$. 
\par
We proceed to derive the repulsive interaction $U_{\rm R}$ from 
the on-shell two-body $T$-matrix with an {\it attractive} 
contact interaction $U$ ($<0$) as~\cite{Ohashi}
\begin{eqnarray}
T(\bm{k},\bm{k};2\varepsilon_{\bm{k}}+i\delta)
&=&\left[\frac{1}{U}-\sum_{\bm{p}}\frac{1}{2\varepsilon_{\bm{k}}+i\delta-2\varepsilon_{\bm{p}}}\right]^{-1}\cr
&=&\frac{4\pi a}{m}\frac{1-ika}{1+ k^2a^2 },
\end{eqnarray}
where we set the effective range to zero by taking an infinitely 
large cutoff $\Lambda$, $\delta$ is a positive infinitesimal, and
$a=\left(\frac{4\pi}{mU}+\frac{2}{\pi}\Lambda\right)^{-1}$ 
is the $s$-wave scattering length.  While the constant repulsive 
interaction $U_{\rm R}=T(\bm{0},\bm{0};0)=\frac{4\pi a}{m}$ 
in the excited branch is usually employed to study 
possible ferromagnetism,  we here keep the incoming 
momentum dependence of the $T$-matrix and obtain
$U_{\rm R}(k)=\frac{4\pi a}{m}\frac{1-ika}{1+ k^2a^2 }$, 
which is complex-valued.  In $U_{\rm R}(k)$ we have set 
$\bm{k}'=\bm{k}$ and $\bm{P}=\bm{0}$ for simplicity.
This complexed-valued effective interaction, which 
behaves as $1/(1+ika)$, reflects the fact that the excited branch is 
unstable against inelastic decay to the molecular state 
of energy $-1/ma^2$ in the attractive branch.  
While this imaginary part is negligible when $a$ is small, 
it would play a significant role near the ferromagnetic transition 
expected to occur in the strongly interacting regime.
\par
{\it Repulsive Fermi polaron}---
To see the feature of our non-Hermitian model, we first 
consider the zero-temperature ($T=0$), highly polarized case 
in which an impurity atom ($\sigma=\dwn$) immersed in a Fermi sea of 
noninteracting majority atoms ($\sigma=\up$) forms a repulsive Fermi polaron.
The retarded Green's function of such an impurity is given by
$G_{\dwn}^{\rm R}(\bm{p},\omega)=\left[\omega-\varepsilon_{\bm{p}}-\Sigma_{\dwn}^{\rm R}(\bm{p},\omega)\right]^{-1}$.
Here,
\begin{eqnarray}
\Sigma_\dwn^{\rm R}(\bm{p})=\sum_{\bm{p}'}U_{\rm R}\left(\frac{|\bm{p}-\bm{p}'|}{2}\right)f(\xi_{\bm{p}',\up}),
\label{selfenergy}
\end{eqnarray}
with the Fermi-Dirac distribution function $f(\omega)=(e^{\beta\omega}+1)^{-1}$,
is the impurity self-energy that retains only the 
lowest-order diagram (Hartree correction). 
This self-energy can be regarded as an approximate form of the 
{\it T}-matrix approximation 
(TMA)~\cite{Combescot,Massignan:2008,MassignanBruun,Hu,Mulkerin,Tajima1,Tajima2}.
While TMA allows for the Pauli blocking of majority atoms 
in intermediate states that are included in the ladder 
diagrams, the present approach neglects such diagrams 
and hence reduces to a mean-field approximation to the medium effect.  
We can obtain the simple analytical expressions for the repulsive 
polaron energy $E_{\rm r}={\rm Re}\Sigma_{\dwn}^{\rm R}(\bm{0})$, 
the decay rate $\Gamma=-2{\rm Im}\Sigma_{\dwn}^{\rm R}(\bm{0})$,
and the effective mass $m/m^*=1+m\left.\frac{\partial^2 {\rm Re}\Sigma_{\dwn}^{\rm R}(\bm{p})}{\partial p^2}\right|_{\bm{p}=\bm{0}}$ 
as 
\begin{eqnarray}
E_{\rm r}&=&\varepsilon_{\rm F,\up}\frac{16}{\pi}\left[\frac{1}{k_{\rm F,\up}a}-\frac{2\tan^{-1}\left(\frac{k_{\rm F,\up}a}{2}\right)}{(k_{\rm F,\up}a)^2}\right],
\end{eqnarray}
\begin{eqnarray}
\Gamma&=&\varepsilon_{\rm F,\up}\frac{32}{\pi}\frac{\left(\frac{k_{\rm F,\up}a}{2}\right)^2-\ln\left[1+\left(\frac{k_{\rm F.\up}a}{2}\right)^2\right]}{(k_{\rm F,\up}a)^2},
\end{eqnarray}
and
\begin{eqnarray}
\frac{m}{m^*}&=&
1-\frac{16}{3\pi}\frac{(k_{\rm F,\up}a)^3}{\left(4+k_{\rm F,\up}^2a^2\right)^2},
\end{eqnarray}
where $\varepsilon_{\rm F,\up}$ and $k_{\rm F,\up}$ are the majority Fermi energy and momentum, respectively.
\par
\begin{figure}[t]
\begin{center}
\includegraphics[width=0.7\hsize]{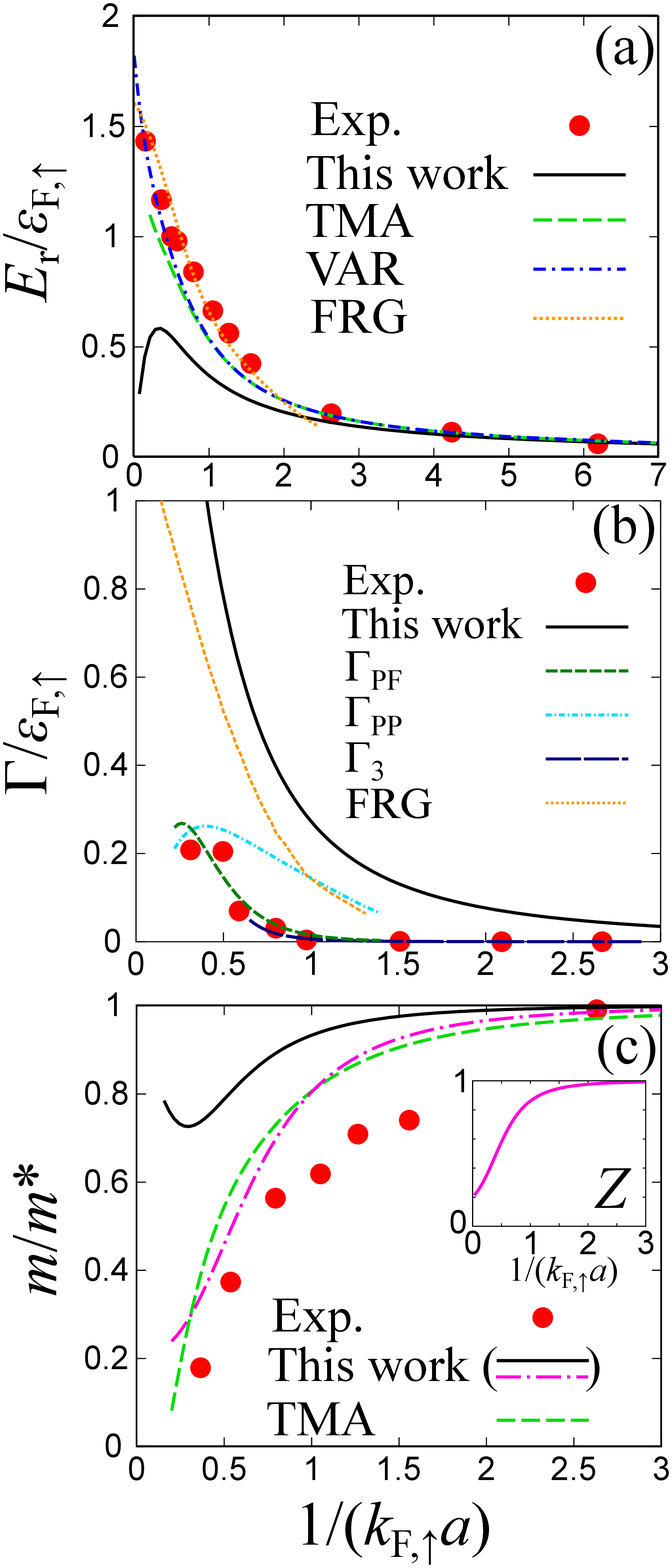}
\end{center}
\caption{(a) Repulsive polaron energy $E_{\rm r}$, (b) decay rate $\Gamma$, 
and (c) effective mass $m^*$, calculated from the non-Hermitian model
as function of $1/k_{\rm F,\up}a$. For comparison, we also show the experimental 
results~\cite{Scazza} and theoretical curves of TMA~\cite{MassignanBruun}, 
variational method (VAR)~\cite{Cui}, and functional renormalization group 
(FRG)~\cite{Schmidt}. In panel (b), the theoretical results for the 
decay rates associated with polaron-to-free particle decay 
($\Gamma_{\rm PF}$)~\cite{Scazza}, polaron-to-polaron decay 
($\Gamma_{\rm PP}$)~\cite{MassignanBruun}, and three-body recombination 
($\Gamma_3$)~\cite{Petrov} are also plotted.
The dash-dotted curve in panel (c) shows $m^*$ modified by the quasi-particle residue $Z$ shown in the inset.}
\label{fig1}
\end{figure}
In Fig.~\ref{fig1} we plot the obtained results 
for (a) $E_{\rm r}$, (b) $\Gamma$, and (c) $m^*$, together
with the experimental results~\cite{Scazza} and earlier 
calculations~\cite{MassignanBruun,Cui,Schmidt,Petrov}.
Although we have used a rather simple method, our result for $E_{\rm r}$ shows a good agreement with the 
experimental and TMA results, particularly in the weakly 
repulsive regime $k_{\rm F,\up}a\lesssim 1$.
This is consistent with the fact that our approach, 
a simplified version of TMA, is what TMA reduces to
in the weak coupling limit.  The overestimation of 
$\Gamma$ in our approach as compared with the empirical 
values even in the weak coupling regime, on the other hand, 
can be understood as lack of the Pauli-blocking effect and of the three-body 
recombination process~\cite{Scazza,Petrov} in the two-body $T$-matrix.  
Incidentally, near the unitarity limit, our results 
underestimate the repulsive polaron energy.
From the comparison between our results and TMA, this may possibly be because our approach ignores
the role played by the Pauli blocking in the attractive branch 
in indirect suppression of the two-body loss.
Fluctuation corrections beyond the present 
mean-field approach may also be responsible for
such underestimate. 
Although the Hartree correction contains no energy dependence 
and hence keeps the quasi-particle residue $Z$ at unity, 
in some cases, even the lowest order correction due to the interaction requires the $Z$ contribution.
For example, 
the effective mass undergoes a leading order modification by $Z$ as $m/m^*=Z\left[1+m\left.\frac{\partial^2 {\rm Re}\Sigma_{\dwn}^{\rm R}(\bm{p})}{\partial p^2}\right|_{\bm{p}=\bm{0}}\right]$, where $Z=\left[1-{\rm Re}\frac{\partial \Sigma_{\dwn}^{\rm R}(\bm{p=0},\omega)}{\partial\omega}\right]^{-1}$
can be estimated by using $U_{\rm R}(\bm{k},\omega_{\up}+\omega_{\dwn})=U_{\rm R}(\bm{k})+{\rm Re}\left[\left.\frac{\partial T(\bm{k},\bm{k},\omega')}{\partial \omega'}\right|_{\omega'=0}\right](\omega_{\up}+\omega_{\dwn})$ with the incoming energy $\omega_{\sigma}$ as
\begin{eqnarray}
\label{eqZ}
Z=\left[1+\frac{8}{\pi}\left[\tan^{-1}\left(\frac{k_{\rm F,\up}a}{2}\right)-\frac{2k_{\rm F,\up}a}{4+k_{\rm F,\up}^2a^2}\right]\right]^{-1}.
\end{eqnarray}
The resultant $m^*$ agrees well with the experimental and TMA results.
\par
{\it Lindblad equation and number densities}---
We next show how the number densities $N_{\sigma}$ 
behaves via the two-body loss suffered by the repulsive 
branch due to the non-Hermitian interaction. 
We assume that the system considered here
is described by an open quantum system in which the 
main system in the repulsive branch interacts with 
the bath in the attractive branch via a 
density-dependent interaction
$\bar{U}_{\rm R}=U_{\rm R}(k=k_{\rm a})$,
where 
$k_{\rm a}=\frac{1}{N_{\up}N_{\dwn}}\sum_{|\bm{k}|\leq k_{\rm F,\up}}\sum_{|\bm{k}'|\leq k_{\rm F,\dwn}}\left|\frac{\bm{k}-\bm{k}'}{2}\right|$
is the averaged relative momentum 
with the Fermi momentum of $\sigma$ atoms $k_{\rm F,\sigma}$.
\par
In describing the quantum dynamics of the system in 
the repulsive branch, we utilize the Lindblad equation 
for the reduced density matrix $\rho$ in the repulsive branch, 
which is given by~\cite{Lindblad,Braaten}
\begin{eqnarray}
i\frac{d\rho}{dt}=[H,\rho]-i\{K,\rho\}+i\int d^3\bm{r}L(\bm{r})\rho L^\dag(\bm{r}),
\label{Lind}
\end{eqnarray}
where $H$ and $K$ are the Hermitian and non-Hermitian parts 
of the effective Hamiltonian (\ref{Heff}), i.e., 
$H_{\rm eff}=H-iK$, and $L(\bm{r})$ is the local Lindblad 
operator as given by
$L(\bm{r})=\sqrt{-2{\rm Im}\bar{U}_{\rm R}}\psi_{\dwn}(\bm{r})\psi_{\up}(\bm{r})$
in such a way as to satisfy
$K=\frac{1}{2}\int d^3\bm{r}L^\dag(\bm{r})L(\bm{r})$.
Using Eq.\ (\ref{Lind}), one can obtain the time derivative of 
$N_{\sigma}={\rm Tr}(\rho\hat{N}_{\sigma})$ with 
the density operator 
$\hat{N}_{\sigma}\equiv
\sum_{\bm{p}}c^{\dagger}_{\bm{p},\sigma}c_{\bm{p},\sigma}$ as
\begin{eqnarray}
\label{eqnt}
\frac{dN_{\sigma}}{dt}
&=&2{\rm Im}\bar{U}_{\rm R}\int d^3\bm{r}\langle\psi_{\up}^\dag(\bm{r})\psi_{\dwn}^\dag(\bm{r})\psi_{\dwn}(\bm{r})\psi_{\up}(\bm{r})\rangle\cr
&\simeq&2{\rm Im}\bar{U}_{\rm R} N_{\up} N_{\dwn},
\end{eqnarray}
where we have used a commutation relation 
$[\hat{N}_{\sigma},L(\bm{r})]=-L(\bm{r})$
as well as the weak coupling approximation in the second line.
From Eq.\ (\ref{eqnt}), one can find 
$\frac{dM}{dt}\equiv\frac{d}{dt}\left(N_{\up}-N_{\dwn}\right)=0$,
which indicates that the two-body loss itself would 
not directly suppress the magnetization $M$, if any at all, 
but would drive such a ferromagnetic system to 
phase separaration between the ferromagnetic 
and molecular states due to decrease of 
the number density in the repulsive branch.
\par
\begin{figure}[t]
\begin{center}
\includegraphics[width=0.7\hsize]{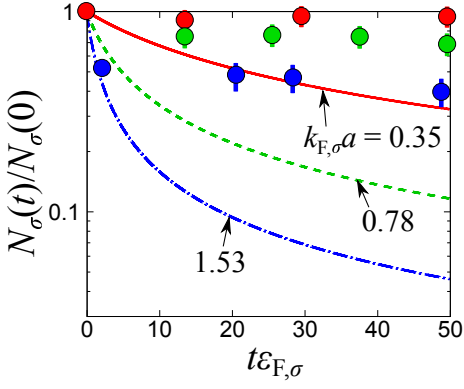}
\end{center}
\caption{Time-dependent atomic number density $N_{\sigma}(t)$ 
obtained from Eq.~(\ref{eqnt}) in an unpolarized Fermi gas 
with $k_{\rm F,\sigma}a=0.35$, $0.78$, and $1.53$.
The circles show the corresponding experimental 
results from Ref.~\cite{Amico}.
}
\label{fig2}
\end{figure}

Figure~\ref{fig2} shows the calculated $N_{\sigma}(t)$ for
an unpolarized repulsive Fermi gas that has 
an initial number density 
$N_{\sigma}(0)=\frac{k_{\rm F,\sigma}^3}{6\pi^2}$, which 
is plotted as function of $t\varepsilon_{\rm F,\sigma}$ 
 with the common Fermi energy
$\varepsilon_{\rm F,\up}=\varepsilon_{\rm F,\dwn}$.
As is consistent with the empirical behavior~\cite{Amico}, 
the resulting decay rate tends to increase with $a$. 
We note, however, that our results overestimate the particle loss 
during the time evolution.
While we here use the density-dependent interaction  
$\bar{U}_{\rm R}$ relevant for the $T=0$ Fermi degeneracy, 
no temperature dependence has been considered.  In the realistic 
case of nonzero temperature, the Fermi distribution has a 
thermal diffuseness, which acts to weaken the interaction by 
allowing scattering between two atoms of lower momenta to occur.
In addition, the inverse process, namely, the pumping to the 
excited branch due to dissociation of a molecule~\cite{Bruun2010} in the 
attractive branch, is neglected in the present analysis.
This process may also contribute because it is required to
understand the observed competition between attractive and 
repulsive correlations~\cite{Amico}.
Recall that as can be seen from the results for 
$\Gamma$ shown in Fig.~\ref{fig1}(b), the three-body process 
is also crucial to reproduction of the empirical behavior,
particularly in the weak coupling regime.
The density inhomogeneity resulting from a harmonic trap as well as strong correlations would also be of importance to describe the experimental result quantitatively.
While more sophisticated treatments are required to describe 
the full time dependence in a quantitative manner,
we restrict ourselves to the present non-Hermitian model
to provide a new insight into 
the many-body physics.
\par
{\it Non-Hermitian ferromagnetism and dynamical instability}---
Finally we discuss effects of the non-Hermitian interaction 
on the ferromagnetic phase transition within the RPA via the susceptibility
\begin{eqnarray}
\label{eqRPA}
\chi^{\rm R}(\bm{q},\omega)=\frac{\chi_0^{\rm R}(\bm{q},\omega)}{1+\bar{U}_{\rm R}\chi_0^{\rm R}(\bm{q},\omega)},
\end{eqnarray}
where $\chi_0^{\rm R}(\bm{q},\omega)$ is the lowest-order 
particle-hole bubble as given by~\cite{Lancaster}
\begin{eqnarray}
\label{eqchi0}
\chi_0^{\rm R}(\bm{q},\omega)&=&\frac{i}{2}\sum_{\bm{k}}\int\frac{d\omega'}{2\pi}\left[G_{\dwn}^{\rm R}(\bm{k}+\bm{q},\omega+\omega')G_{\up}^{\rm K}(\bm{k},\omega')\right.\cr
&&\left.+G_{\dwn}^{\rm K}(\bm{k}+\bm{q},\omega+\omega')G_{\up}^{\rm A}(\bm{k},\omega')\right].
\end{eqnarray}
In Eq.\ (\ref{eqchi0}), $G^{\rm R,A,K}$ are the retarded, advanced, 
and Keldysh Green's functions~\cite{Rammer}.  For simplicity, 
we assume that 
the time-dependence of the number density in the metastable state 
is sufficiently slow to use the Keldysh component 
in thermal equilibrium as
$G_{\sigma}^{\rm K}(\bm{k},\omega)\simeq[G_{\sigma}^{\rm R}(\bm{k},\omega)-G_{\sigma}^{\rm A}(\bm{k},\omega)]\tanh\left(\frac{\omega}{2T}\right)$~\cite{Rammer}.
We then determine the chemical potential $\mu_{\sigma}$ 
by fixing the number density given by
\begin{eqnarray}
\label{eqN}
N_{\sigma}
&=&-\frac{1}{\pi}\sum_{\bm{p}}\int d\omega f(\omega){\rm Im}G_{\sigma}^{\rm R}(\bm{p},\omega).
\end{eqnarray}
While $\mu_{\sigma}$ is real-valued in thermal equilibrium,
it is not necessarily so in a non-equilibrium state, 
as pointed out in Ref.~\cite{Iskin}.
In fact, the chemical potential, generally defined 
as the internal energy difference when another particle is 
added to the system, i.e., 
$\mu_{\sigma}=\frac{\partial E}{\partial N_\sigma}$,
reads 
$\mu_{\sigma}=\varepsilon_{\rm F,\sigma}+\bar{U}_{\rm R}N_{-\sigma}\in \mathbb{C}$.
This is because within the mean-field approximation, 
$ E = \frac{3}{5}N_{\up}\varepsilon_{\rm F,\up}+\frac{3}{5}N_{\dwn}\varepsilon_{\rm F,\dwn}+\bar{U}_{\rm R}N_{\up}N_{\dwn}\in \mathbb{C}$.
By substituting this $\mu_{\sigma}$ to $G_{\sigma}^{\rm R}(\bm{p},\omega)=\left(\omega-\varepsilon_{\bm{p}}+\mu_{\sigma}-\bar{U}_{\rm R}N_{-\sigma}+i\delta\right)^{-1}$, 
one can find that ${\rm Im}\bar{U}_{\rm R}N_{-\sigma}$ is compensated 
by the imaginary part of $\mu_{\sigma}$.
Consequently, $\chi_0^{\rm R}(\bm{q},\omega)$ reduces to the well-known 
Lindhard function,
\begin{eqnarray}
\chi_0^{\rm R}(\bm{q},\omega)=-\sum_{\bm{k}}
\frac{f(\xi_{\bm{k+q},\dwn}^*)-f(\xi_{\bm{k},\up}^*)}{\omega+\xi_{\bm{k},\up}^*-\xi_{\bm{k+q},\dwn}^*+i\delta},
\end{eqnarray}
where $\xi_{\bm{k},\sigma}^*=\varepsilon_{\bm{k}}-\mu_{\sigma}^*$ with the renormalized chemical potential 
$\mu_{\sigma}^{*}=\mu_{\sigma}-\bar{U}_{\rm R}N_{-\sigma}$ (corresponding to the Fermi energy $\varepsilon_{\rm F,\sigma}$) is the excitation energy. 
It is to be noted that $\mu_{\sigma}^{*}$ is now real-valued, a feature that is also 
consistent with the number equation (\ref{eqN}).
\par
Finally, we address what kind of ferromagnetic state is realized in the presence of the inelastic two-body decay.  
To this end, for simplicity, we focus on the 
zero-momentum pole $\Omega$ of the RPA susceptibility 
$\chi^{\rm R}(\bm{q},\omega)$, Eq.\ (\ref{eqRPA}), which 
fulfills
\begin{eqnarray}
1+\bar{U}_{\rm R}\chi_0^{\rm R}(\bm{q}\rightarrow\bm{0},\Omega)&=&0.
\label{stoner}
\end{eqnarray}
In the unpolarized case with $\Omega=0$, the critical 
density that satisfies Eq.\ (\ref{stoner}) corresponds to 
the condition for the Stoner instability, i.e., 
spontaneous polarization occurs statically and uniformly.
In the presence of such polarization along which
we take the direction of $\up$, we obtain a solution to
Eq.\ (\ref{stoner}) as~\cite{Note}
\begin{eqnarray}
\label{eq15}
{\rm Re}\Omega=\mu_{\up}^*-\mu_{\dwn}^*-{\rm Re}\bar{U}_{\rm R}(N_{\up}-N_{\dwn}),
\end{eqnarray}
\begin{eqnarray}
\label{eq16}
{\rm Im}\Omega=-{\rm Im}\bar{U}_{\rm R}(N_{\up}-N_{\dwn})>0.
\end{eqnarray}
The pole in the upper complex plane of frequency indicates 
that the system undergoes a dynamical instability 
once the system has nonzero polarization.
This is because the inverse Fourier transformation 
of $\chi_0^{\rm R}(\bm{0},\omega)$ with respect to 
frequency leads to the time dependence like 
$\sim \exp(-i\Omega t)=\exp(-i{\rm Re}\Omega t+{\rm Im}\Omega t)$.
This exponential growth of spin fluctuations,
together with the above argument based on 
Eq.\ (\ref{eqnt}), suggests
that the ferromagnetic state does not undergo a phase
transition back to the homogeneous paramagnetic 
(unpolarized) phase, but to a qualitatively 
different state that is reminiscent of a heterogeneous 
phase observed in recent experiments~\cite{Scazza:2020}.
We remark in passing that within the present
model, a homogeneous transition from the 
paramagnetic to ferromagnetic phase with increasing 
density is unlikely to occur since the inelastic decay
to the molecular state is designed to keep decreasing 
the density without any feedback.  Recall that this caveat 
holds also for the study of the repulsive polaron energy 
as well as the decay of the atom number density.
Further investigations beyond the present framework 
would be desired.
\par
{\it Summary}---
In this work, we have developed a non-Hermitian effective theory 
to describe the polaronic and magnetic properties of
a repulsively interacting Fermi gas in the excited branch.
This theory incorporates the complex-valued interaction 
obtained from the on-shell two-body $T$-matrix in such 
a way as to characterize the two-body inelastic decay to the 
molecular state in the attractive branch.  Within the weak coupling 
approximation, we have derived simple analytical formulas 
for the repulsive polaron properties and the differential equation 
for the time-dependent atomic number density, which explain
the experimental results fairly well given the overestimated
two-body inelastic loss.
By building the present non-Hermitian framework into 
the analysis of the dynamical transverse spin susceptibility 
within the RPA, we show that the 
complex-valued two-body interaction drives a 
uniform ferromagnetic system unstable to
heterogenuity as observed in recent experiments. 
To investigate the spatial scale of the heterogeneous phase,
the momentum dependence of the time-dependent susceptibility would have to be clarified.
\par 

\par
The authors thank F. Scazza for providing us with their data 
in Ref.~\cite{Scazza} and E. Nakano, J. Takahashi, K. Nishimura, 
T. Hata, T. M. Doi, and S. Tsutsui for useful discussions. 
This work is supported by Grants-in-Aid for JSPS fellows (No.\ 17J03975) and 
for Scientific Research from JSPS (Nos.\ 18H01211 and 18H05406).


\begin{thebibliography}{99}
\bibitem{Note} The solution of Eq.~(\ref{stoner}) at nonzero 
${\rm Re}\Omega$ would mean the presence of a gapped 
collective mode (magnon) in the system with a broken 
spin-inversion symmetry.  Then, the solution at vanishing 
${\rm Re}\Omega$ would give a condition at which the magnon 
becomes gapless and hence domains of itinerant ferromagnetism 
occur.  Once an impurity atom ($\dwn$) is put into a resultant  
magnetic domain ($\up$), the repulsive polaron formed
has an energy $E_{\rm r}=\varepsilon_{\rm F,\up}$, which can 
be obtained by substituting into Eq.\ (\ref{eq15}) the relevant
conditions ${\rm Re}\Omega=0$, $N_\dwn=0$, and $\mu_\dwn^*=0$
as well as using Eq.\ (\ref{selfenergy}).
This just corresponds to the critical stability condition of
the configuration that consists of magnetic domains 
as given in Ref.~\cite{MassignanBruun}.
\end{thebibliography}
\end{document}